\documentclass[12pt,preprint]{aastex}

\newcommand{\mgi}{Mg\,I}
\newcommand{\mgii}{Mg\,II}
\newcommand{\invcm}{cm$^{-1}$}

\shorttitle{Mg\,I emission from Procyon}
\shortauthors{Ryde \&\ Richter}

\begin{document}


\title{Non-thermal \mgi\ emission at 12 \micron\ from Procyon}


\author{N. Ryde\altaffilmark{1,2}}
\affil{Department of Astronomy and Space Physics, Uppsala University, Box 515, SE-75120 Uppsala, Sweden}
\email{ryde@astro.uu.se}

\author{M. J. Richter\altaffilmark{2}}
\affil{Department of Physics, University of California at Davis, CA 95616}
\email{richter@physics.ucdavis.edu}


\altaffiltext{1}{Postdoctoral Fellow in 2001-2002 at the Department of Astronomy, University of Texas at Austin, TX 78712}
\altaffiltext{2}{Visiting Astronomer at the Infrared Telescope Facility,
which is operated by the University of Hawaii under Cooperative Agreement
no. NCC 5-538 with the National Aeronautics and Space Administration, Office
of Space Science, Planetary Astronomy Program.}
%\altaffiltext{2}{Society of Fellows, Harvard University.}
%\altaffiltext{3}{present address: Center for Astrophysics,
%    60 Garden Street, Cambridge, MA 02138}
%\altaffiltext{4}{Visiting Programmer, Space Telescope Science Institute}
%\altaffiltext{5}{Patron, Alonso's Bar and Grill}


\begin{abstract}
We report on stellar \mgi\ emission at 12\,\micron\ from
$\alpha$ CMi (Procyon), a star slightly hotter than the Sun.
Solar \mgi\ emission is well-known and its formation
was successfully explained
% for the first time
in detail by \cite{mc_mg}.  Here, for the first time, we compare
synthetic spectra of the emission lines at $12\,\micron$ with
observations of a star other than the Sun.
The use of these lines as stellar diagnostics
has been anticipated for 10 years or more \cite[see, e.g., ][]{mc_mg}.
We find that the model reproduces the observed emission in Procyon
quite well.  We expect that high-resolution spectrographs on
$8-10$ m telescopes will finally be able to exploit these new diagnostics.

\end{abstract}


%% subject header.  Objects should be in the appropriate "individual"
%% headers (e.g. quasars: individual, stars: individual, etc.) with the
%% additional provision that the total number of headers, including each
%% individual object, not exceed six.  The \objectname{} macro, and its
%% alias \object{}, is used to mark each object.  The macro takes the object
%% name as its primary argument.  This name will appear in the paper
%% and serve as the link's anchor in the electronic edition if the name
%% is recognized by the data centers.  The macro also takes an optional
%% argument in parentheses in cases where the data center identification
%% differs from what is to be printed in the paper.



\section{Introduction}

The origin and formation process of the once enigmatic \emph{solar} \mgi\
emission at 811.578 \invcm\ (12.3217~\micron) and
818.058~\invcm\ (12.2241~\micron) was first explained successfully and
in detail by \citet[hereafter CRS]{mc_mg}. By employing
an extensive
model atom of \mgi\ in a full
Non Local Thermodynamic Equilibrium (NLTE) calculation,
they concluded the emission lines were photospheric in origin.
CRS accounted excellently
for the emission strengths of the \mgi\ lines,
their complicated intensity profiles as a function of viewing angle
on the solar disk, and the relative
strengths of several \mgi\ lines, both in emission and absorption.
%and can, therefore, not be used as a probe of the chromosphere.
Furthermore, CRS could explain
the similar but weaker Si\,{I} and Al\,{I} emission features observed in the solar spectrum.
\cite{baumuller} successfully modeled
%the above mentioned
these Al\,{I} emission-features in the Sun, showing that a similar
NLTE effect as for \mgi\ is responsible.
\cite{zhao} were also able to reproduce the solar emission lines with a similar NLTE analysis of
magnesium in the solar photosphere. Moreover, by including artificially reduced,
inelastic, neutral-hydrogen collisions,
%(which should be
%increasingly important for cool, metal-poor stars),
they were also able to simultaneously model line cores of visual and near-infrared (NIR)
\mgi\ absorption-lines in the range of
%$4571-11\,828$\,\AA.
$4\,500-12\,000$\,\AA.

%%% Section added by mjr:

The \mgi\ emission lines at 12~\micron\ in the Sun clearly show Zeeman
splitting and so trace the local magnetic field~\citep{brault83,bruls:2}.
Because Zeeman splitting
depends on the square of the wavelength while Doppler broadening goes
as the first power, mid-infrared lines better reveal Zeeman splitting
than optical or NIR lines with the same Land\'{e} $g$-factor.  Thus, these lines
potentially offer an excellent method
for measuring stellar disk-averaged magnetic fields, and will be
complementary to studies using NIR lines in that they will probe other atmospheric layers and
are more Zeeman sensitive.

%%%%

In this paper we report on our detection and modeling of the mid-infrared
\mgi\ emission features
in Procyon, an F5IV-V star.
In the Sun, the observed emission lines are formed as a result of a NLTE
flow cycle.   As lines originating from high-excitation levels
become optically thin in the photosphere, lower lying levels are
overpopulated and subsequently photoionized.
The \mgii\ reservoir refills the high-excitation levels via Rydberg states of \mgi, leading to the emission.
%In our solar model-atmosphere, more than 93\% of the magnesium is
%in \mgii\ all through the line-forming regions while for Procyon,
%more than 99\% of the magnesium
%is in Mg\,{Ii}.
From NLTE calculations, the relevant departures from Boltzmann
level-populations
for the levels involved --
$3s7i^{1,3}\mathrm{I}^e$  and $3s6h^{1,3}\mathrm{H}^o$ for the
811.6 \invcm\ line and
$3s7h^{1,3}\mathrm{H}^o$ and $3s6g^{1,3}\mathrm{G}^e$ for the 818.1
\invcm\ line --
are quite similar for
both the upper and lower states. The departure coefficients of
the upper levels are of the order of 10\% larger than those of the lower states.
However, this small difference between the level departures
is the direct cause of the observed emission.

\section{Observations}

We observed Procyon with the Texas Echelon-Cross-Echelle Spectrograph~\citep[TEXES, ][]{lacy02},
a visitor instrument at the 3m NASA Infrared Telescope Facility (IRTF).
The raw data were collected during
November 2000, November 2001, December 2002, December 2003, and
January 2004.
Separate grating settings were required for each \mgi\ line.

We used the cross-dispersed, high spectral resolution mode for all
observations.  Observations
of unresolved line sources indicate
that the core of the line profile at this wavelength
is reasonably reproduced
with a Gaussian with FWHM $=3.5$~km\,s$^{-1}$ ($R\sim 86\,000$) and wings that are
slightly broader than Gaussian.

Data acquisition and reduction followed the standard procedure described
in~\citet{lacy02}.
To remove sky and telescope background,
Procyon was nodded along the slit.
Wavelength calibration was done using telluric
atmospheric lines; previous experience shows this procedure typically
gives accuracy better than 1~km\,s$^{-1}$.
After reducing a given data set, we normalized the continuum
using a 6$^\mathrm{th}$ order Legendre polynomial and then
combined appropriate data files.
Procyon has almost no photospheric features at this wavelength so
determining the continuum is easy and reliable.  Furthermore,
the telluric atmosphere at the \mgi\ wavelengths is very clean.
Slight variations in wavelength setting from night-to-night result in
increased noise where less data could be coadded.

%% In a manner similar to \objectname authors can provide links to dataset
%% hosted at participating data centers via the \dataset{} command.  The
%% second curly bracket argument is printed in the text while the first
%% parentheses argument serves as the valid data set identifier.  Large
%% lists of data set are best provided in a table (see Table 3 for an example).
%% Valid data set identifiers should be obtained from the data center that
%% is currently hosting the data.

%% In this section, we use  the \subsection command to set off
%% a subsection.  \footnote is used to insert a footnote to the text.


%% This section also includes several of the displayed math environments
%% mentioned in the Author Guide.

%\section{Analysis}

\section{Modeling}

%In this section we first describe the model used and
%then discuss the results.

%\subsection{The model}

We have analyzed the Procyon \mgi\ lines
as CRS successfully did for the \mgi\ lines in the Sun.
We use a one-dimensional, hydrostatic, flux-conserved, non-magnetic, LTE
model for the physical structure of the atmosphere.
The model-atmosphere, including the specific, mean-intensity field at
all depths,
is calculated with the {\sc marcs} code, which was first developed by \cite{marcs} and has been
successively updated ever since (Gustafsson et al. 2004, in preparation).

For the line formation of \mgi\ and the spectral
synthesis, a full NLTE calculation is performed using
the program MULTI \citep{multi}.
Our \mgi\ model atom was kindly provided by Mats Carlsson.
We require our model to reproduce the excellent fits of the
intensity profiles measured at different locations on the solar
disk as presented in CRS.
%{\it\large  Does this mean you reproduced
%the fits presented in CRS?}yes
We solve the equations of statistical equilibrium for the magnesium
atom level populations at each of 66 levels, including 315
line transitions and 65 bound-free transitions (ionization
and recombination).
One \mgii\ level is included.
We calculate photoionization rates
%using a photospheric radiation field that
by incorporating the calculated specific, mean-intensity
field for all depths from the model atmosphere.
This treatment allows the full line-blanketing to be considered through
ultraviolet wavelengths, which mainly affect the
photoionization from the lowest states.
For a further discussion, references for the atomic data,
and the uncertainties in the modeled emission, see CRS.

%zzz
%Collisions

The fundamental stellar parameters of Procyon, needed in
the calculation of the model atmosphere,
are relatively well known.  All parameters, except the metallicity,
can be derived from direct measurements.
Recent works determining Procyon's parameters include,
\citet{fuhrmann}, \citet{carlos}, and \cite{korn}.
The parameters we have used are:
 an effective temperature of $T_\mathrm{eff} = (6\,512\pm 50)$~K,
a surface gravity of $\log g = 3.96\pm0.02$ (cgs units),
a mass of $M=(1.42\pm0.06)\,M_\odot$,
and a radius of $R=(2.07\pm0.02)\,R_\odot$~\citep{carlos}.
The metallicity (iron abundance)
of the star is slightly lower than solar.
\citet{fuhrmann} measures a slightly super-solar $\alpha$-element abundance.
Nevertheless, we will assume a solar abundance mixture as given by \citet{solar}.
We model the atmosphere with a depth-independent micro-turbulence parameter
of $\xi_\mathrm{micro}=2.0$ km\,s$^{-1}$, in agreement with current literature.

%\subsection{Results and discussion}
\section{Results and discussion}

\begin{figure}
\epsscale{1.00}
\plotone{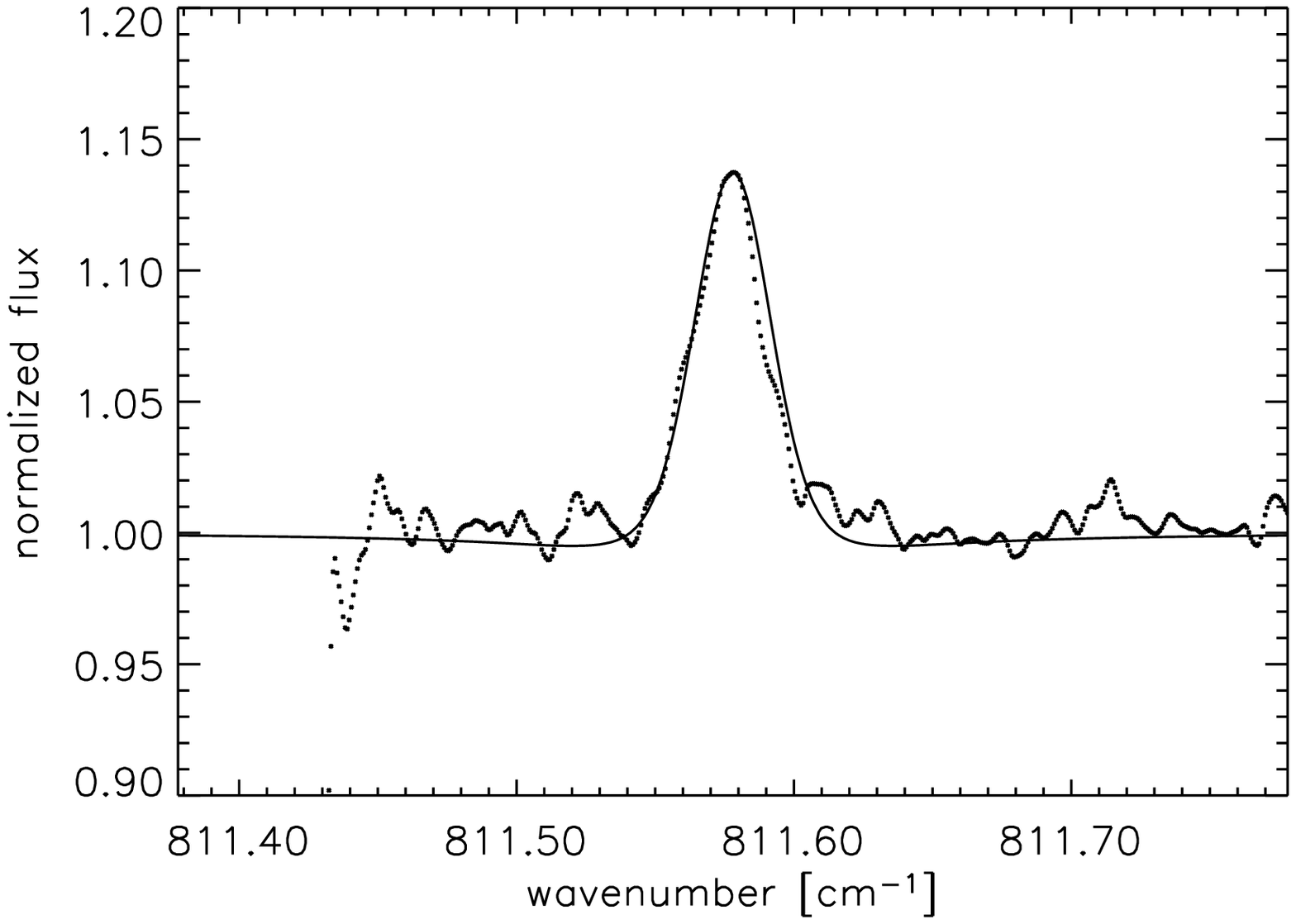}
\caption{Dotted line shows
the \mgi\ emission-line at 811.578 \invcm\ ($12.3217\,\micron$)
observed from Procyon and
the full line shows the model.
 \label{em_811}}
\end{figure}

\begin{figure}
\epsscale{1.00}
\plotone{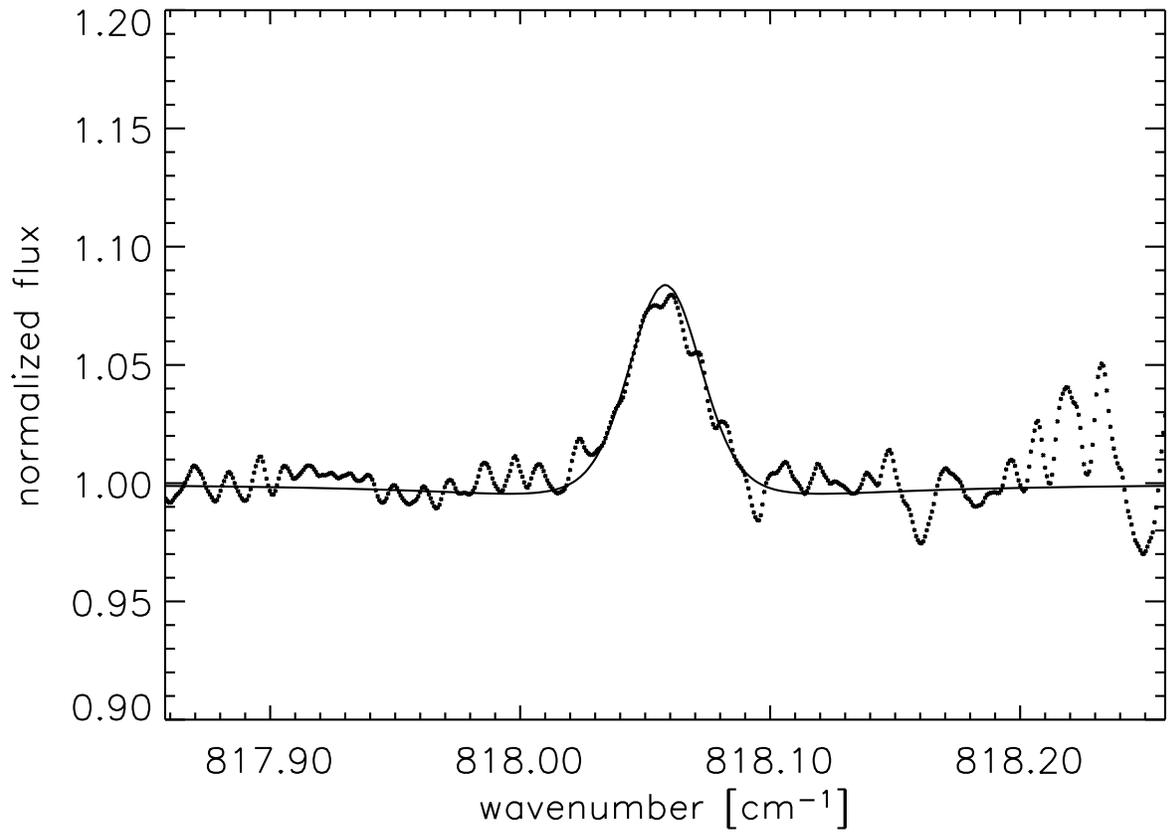}
\caption{Dotted line depicts the 818.058 \invcm\ ($12.2241\,\micron$)
\mgi\ line observed from Procyon.
As in Figure 1, the model prediction is shown by a full line.
\label{em_818}}
\end{figure}

%Given the success in explaining the formation mechanism of the
%solar \mgi\ lines, it is of interest to confront the model also with
%a stellar case.
Our model and observations are shown in Figure~\ref{em_811}
and \ref{em_818}.
In order to match the line widths, we introduce the customary
artifice of macroturbulent broadening, including
the instrumental profile.
We assume the broadening velocities have a
Gaussian distribution with a FWHM of $9.5\,\mathrm{km\,s^{-1}}$, which
is slightly larger than that
given in \citet{carlos}.
We find the model reproduces the disk-averaged observations within the modeling uncertainties.
In a forthcoming paper
%(Ryde et al. 2005, in preparation),
we will
investigate our calculation by
discussing the atomic model of \mgi\
and the assumptions made in our NLTE calculation,
such as the collisional rates, atmospheric radiation field
used and the validity of the assumption of atmospheric homogeneity.

\begin{figure}
\epsscale{1.00}
\plotone{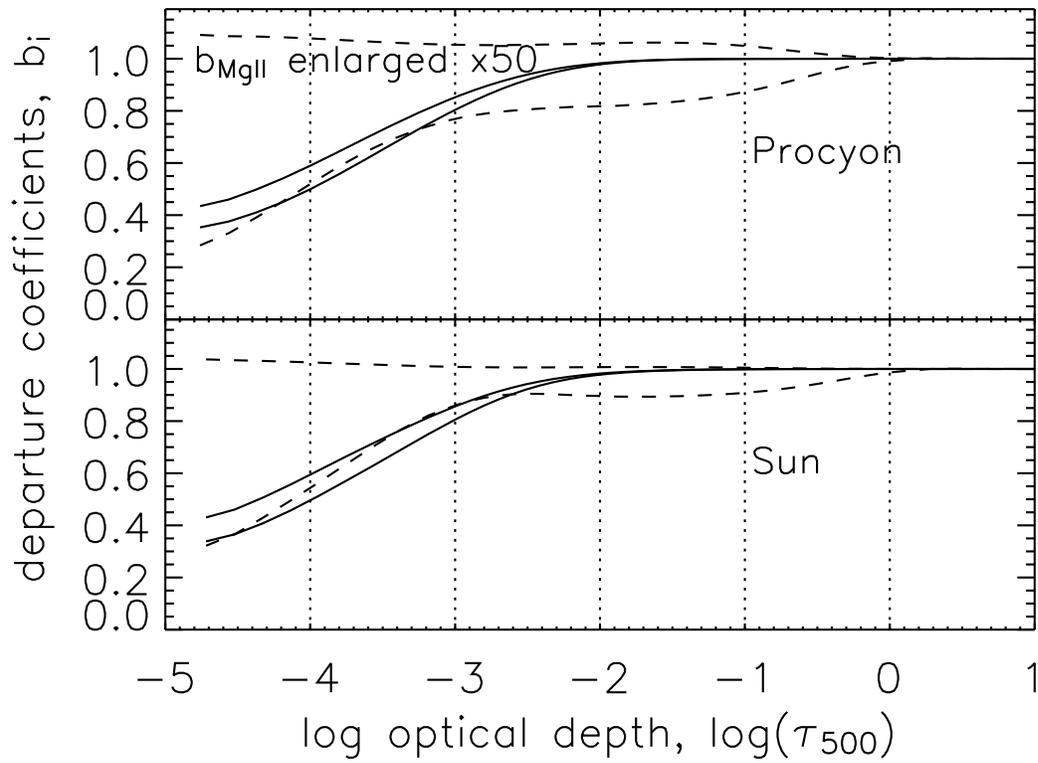}
\caption{Departure coefficients
for the levels involved in the two emission lines are shown
%the 811.6 and 818.1 \invcm\ lines are shown
with full lines.
% for Procyon and the Sun.
%The lower and upper levels of the two lines behave similarly, although
%the lower levels depart more from LTE.
The upper dashed line shows the departure coefficient of \mgii\ and
%, showing a larger magnitude for the Sun.
the lower dashed line shows the ground level of \mgi.
\label{dep_coeff}}
\end{figure}

In Figure~\ref{dep_coeff}, the departure coefficients (defined as in
CRS) of the relevant levels
are plotted as a function of depth in the atmosphere,
shown as the optical depth calculated in the continuum at 500~nm.
The departure coefficients are very similar to the solar case,
indicating that the same formation process of
the \mgi\ emission lines
is at play for Procyon.
Following CRS, we have
included `quasi-elastic \emph{l}-changing'\footnote
{\emph{l} being the azimuthal quantum number}
collisions with neutral particles, which
keep all close-lying Rydberg states with common principal quantum-numbers \emph{n} in
relative thermalization. This means that the departure
coefficients of the upper and lower states, respectively, of the
two \mgi\ emission lines
%($3s7i^{1,3}\mathrm{I}^e \rightarrow 3s6h^{1,3}\mathrm{H}^o$
%and $3s7h^{1,3}\mathrm{H}^o \rightarrow 3s6g^{1,3}\mathrm{G}^e$) will
will follow each other exactly
since they have the same \emph{n} quantum numbers.
Thus, in Figure~\ref{dep_coeff} the upper and lower levels
of the two \mgi\ emission lines fall on top of each other.
The lower levels depart more from LTE.
Whereas for the Sun the population of \mgii\ is slightly larger than that of LTE, for Procyon
the \mgii\ population is very close to LTE.
%In Figure~\ref{dep_coeff}, the departure coefficient of \mgii in Procyon is
%enlarged by a factor of 50.

The \mgi\ emission in Procyon is only marginally larger
than that of the Sun.
This fact can qualitatively be understood as follows.
The higher temperatures in Procyon compared to the Sun,
will lead to a larger ionization of magnesium,  with a factor of ten
less \mgi\ expected all through the line-forming regions.
However, in the line-forming regions of Procyon, the 500~K warmer
temperatures compared to the Sun, also lead to
roughly a factor of ten more atoms
excited to the levels involved in the formation of
the emission line.
Since the departure coefficients for both Procyon and the Sun
are quite similar,
the resulting emission will be of the same order of magnitude,
as observed.

%{\it\large  I took out "first" detection since we included a plot
%of Arcturus with \mgi\ emission in our PASP paper.}
To conclude, we have shown that,
with a NLTE calculation using the CRS model atom of magnesium, it is possible to reproduce \mgi\ emission lines
at 12~\micron\ successfully,
not only for the Sun but also for Procyon. This is the first star, other than the Sun, that
this has been done for. Future investigations of different types of stars, showing \mgi\ emission lines,
will also be important.
%With a solid understanding of their formation process,
The \mgi\ emission lines should be useful tools for measuring stellar
magnetic-fields through their Zeeman splitting. After a decade of anticipation,
high-resolution, mid-IR spectrographs, such as TEXES,
at 8-10 m telescopes will now allow the use of the \mgi\ $12\,\micron$
lines for stellar magnetic-field diagnostics.

\acknowledgments

We thank M. Carlsson for providing the \mgi\ model-atom used in
this study. K. Eriksson, B. Gustafsson, D. Kiselman, A. Korn, and the referee
are acknowledged for fruitful discussions. We are grateful for the help of J. Lacy and T. Greathouse of the {\sc
texes}
team as well as the {\sc irtf} staff.
This work was
supported in part by the Swedish Research Councils {\sc vr} and {\sc stint}
%, the Swedish Foundation for International Cooperation
%in Research and Higher Education,
%Stiftelsen Blanceflor
%Boncompagni-Ludovisi, n\'ee Bildt,
and {\sc nsf} grant AST-0307497.
The construction of {\sc texes} was
supported by grants from the {\sc nsf} and observing with {\sc
texes} was supported by the Texas Advanced Research Program.

%\bibliography{aamnem99,Ref4,KPNO,Ref5,Ref_FL,Ref_posters,ref,ref2,ref3,ref_ryde,Ref_Rdor}
%\bibliographystyle{aabib99}

\clearpage

\end{document}